\documentclass[twocolumn,showpacs,preprintnumbers,amsmath,amssymb]{revtex4}
\usepackage{amssymb}
\usepackage[dvips]{graphicx}
\begin{document}
\draft
\title{Triplet pairing and upper critical field in the mixed state of d-wave
superconductors}
\author{V. V. Kabanov}
\affiliation {\it Josef Stefan Institute 1001, Ljubljana,
Slovenia}

\begin{abstract}
We show that an additional triplet component of the order parameter
is generated in the vortex phase of the d-wave superconductor.
Spatial variations of the triplet component are analyzed for a
strong spin-orbit coupling. Corrections to the London equation and
an unusual temperature dependence of the upper critical field,
$H_{c2}(T)$, are obtained in the case of the weak spin-orbit
coupling.
\end{abstract}
\maketitle
There is a common belief that the copper-based high-temperature
superconductors have nontrivial order parameter (OP) transforming
as $B_{1g}(x^{2}-y^{2})$ representation of $D_{4h}$ point group
\cite{tsuei}. In the framework of the BCS theory this "d-wave"
symmetry is related to the 'internal' coordinate of the OP,
${\mathbf{r}}={\mathbf{r}}_{1}-{\mathbf{r}}_{2}$, assuming its
decomposition as $\Delta({\mathbf{r}}_{1},{\mathbf{r}}_{2})=
\Delta({\mathbf{R}}) \Delta({\mathbf{r}})$, where  ${\mathbf{R}}=
({\mathbf{r}}_{1}+ {\mathbf{r}}_{2})/2$ is the center-of-mass
coordinate \cite{ale}. This symmetry of the OP implies that the
spectrum of elementary excitations is gapless in certain
directions so that the d-wave gap is very sensitive to an external
perturbation. For example, applying a uniform magnetic field
generates  a secondary singlet component of the OP, that lowers
the symmetry and opens a full gap in the nodal directions
\cite{balatsky,kt}. P-wave component of the OP  could be also
generated by a surface induced spin-orbit coupling (SOC)
\cite{gr}. Similarly, superconducting currents give rise to $ip$
secondary component and partial opening of the gap in the nodal
directions \cite{kab}. Appearance of $ip$ component \cite{kab} is
associated with the Lifshitz invariants (LI) in the free energy
(FE) \cite{minsam}. More recently it has been shown that the
triplet OP $i\Delta_{t}$ is generated in the mixed state of
type-II superconductors \cite{lebed}. It was argued that this
effect is a consequence of broken symmetry in the spin space due
to the paramagnetic effect and a broken translational invariance.
This effect was called as type-IV supeconductivity \cite{lebed}.

Phenomenologically nontrivial secondary OP is generated by linear
coupling of the primary OP gradients to the secondary OP. From the
microscopic point of view the interaction of electrons on the
Fermi surface is repulsive in the channel of secondary OP.
Nevertheless, non-diagonal terms corresponding to the coupling to
the primary OP leads to nonzero value of the secondary
OP\cite{lebed,kab}. The relative amplitude of the p-wave component
depends on the strength of SOC and may be of the order
10-20\%\cite{kab}.

Previously it was shown that  the superconducting current in the
d-wave superconductor reduces the symmetry group and generates the
secondary triplet OP \cite{kab}. It is possible since the symmetry
allows for the presence of  terms in the Ginzburg-Landau FE
functional which are of the first order in gradients. Usually
these terms are allowed for crystals where the inversion symmetry
is broken. Inversion is the symmetry operation of $D_{4h}$ group
and  generation of the LI is possible if the secondary component
breaks the inversion symmetry \cite{minsam,levan}. Here we
investigate the electromagnetic response in the mixed state of the
type-II d-wave superconductor in the presence of LI
\cite{minsam,kab} and formulate the criteria for experimental
observation of the effect.

First let us briefly discuss the symmetry properties of  p-wave
superconducting OP \cite{sammin}. The triplet OP has 9 components.
In the case of the strong SOC spin is coupled to the lattice and
transforms together with the lattice. P-wave component of $D_{4h}$
point group is written as
$\mathbf{d}(\mathbf{k,R})=(p_{x}(\mathbf{R})k_{x}+
p_{y}(\mathbf{R})k_{y})\mathbf{\hat{z}}$, where $\mathbf{\hat{z}}$
is a unite axial vector in the spin space and $p_{x}(\mathbf{R})$
and $p_{y}(\mathbf{R})$ plays the role of the OP in the
Ginzburg-Landau functional. In the case of the weak SOC the spin
and the orbital parts of the OP transforms independently and
$\mathbf{d}(\mathbf{k,R})=\mathbf{p}_{x}(\mathbf{R})k_{x}+
\mathbf{p}_{y}(\mathbf{R})k_{y}$, where two vectors in the spin
space $\mathbf{p}_{x}(\mathbf{R}),\mathbf{p}_{y}(\mathbf{R})$ play
the role of the OP. In the latter case the coupling between
singlet and triplet OP is possible only in the presence of the
magnetic field.

$Strong$ $SOC$. The FE of d-wave superconductor per unite of
volume can be written as:
\begin{eqnarray}
F &=& \frac{\hbar^{2}}{2m}|{\mathbf{D}}\psi|^{2}+\alpha(T-T_{c})
|\psi|^{2}+\frac{\beta}{2}|\psi|^{4}+\nonumber\\& & i\eta\Bigl
[\psi^{*} (D_{x}p_{x}-D_{y}p_{y})-\psi
(D_{x}^{*}p_{x}^{*}-D_{y}^{*}p_{y}^{*})\Bigr ]\\& & +\alpha_{p}
(|p_{x}|^{2}+|p_{y}|^{2})+\frac{\mathbf{H}^{2}}{8\pi}\nonumber
\end{eqnarray}
where $\psi$ and $p_{x,y}$ are the primary d-wave and the
secondary p-wave OP, ${\mathbf{D}}=-i\nabla-2e{\mathbf{A}}$,
${\mathbf{A}}$ is the vector potential for magnetic field
${\mathbf{H}}=\nabla\times\mathbf{A}$ , $\alpha,\alpha_{p},\beta
>0, \eta$ are real constants. Here we write explicitly the LI
describing   $p$-wave component. In Eq.(1) the additional
term,  quadratic in $p_{x}$, $p_{y}$,is written to secure
$p_{x}=p_{y}=0$ solution in the uniform state $\psi_{d}=const$.

Minimizing the FE for variation of $\delta \psi,\delta
p_{x},\delta p_{y}$ we obtain the following set of equations:
\begin{eqnarray}
\frac{\hbar^{2}}{2m}{\mathbf{D}}^{2}\psi+\alpha(T-T_{c})\psi +
\beta |\psi|^{2}\psi + \nonumber \\i\eta(D_{x}p_{x}-D_{y}p_{y})=0
\end{eqnarray}
\begin{eqnarray}
p_{x} = \frac{i\eta}{\alpha_{p}}D_{x}\psi \nonumber \\
p_{y} = \frac{-i\eta}{\alpha_{p}}D_{y}\psi
\end{eqnarray}
Substituting Eq.(3) to Eq.(2) we obtain standard Ginzburg-Landau
equation
\begin{equation}
\frac{\hbar^{2}}{2m^{*}}{\mathbf{D}}^{2}\psi+\alpha(T-T_{c})\psi+
\beta |\psi|^{2}\psi=0
\end{equation}
with renormalized effective mass
$(m^{*})^{-1}=m^{-1}-2\eta^2/\alpha_{p}$. The renormalization of
the effective mass was derived in Eq.(7) of  Ref.\cite{kab}. If
renormalized effective mass is negative $(m^{*})^{-1} < 0$
($2\eta^{2}/\alpha_{p} >m^{-1}$) the helical phase is
formed\cite{minsam} and higher order gradient terms for secondary
OP should be included to the FE, Eq.(1). The properties and the
thermodynamics of helical phases have been studied in
Ref.\cite{minsam} and are not considered here. If renormalized
effective mass $(m^{*})^{-1}
>0$ ($2\eta^{2}/\alpha_{p} <m^{-1}$) the FE, Eq.(1),
describes the formation of the secondary OP in the presence of the
magnetic field or in the presence of the supercurrent \cite{kab}.
Therefore we assume in the following that $(m^{*})^{-1}>0$ and
higher order gradient terms are not essential in generation of the
secondary OP.

Minimizing the FE for variations of the vector potential $\delta
{\mathbf{A}}$ and substituting Eqs.(3,4) we obtain superconducting
current:
\begin{equation}
{\mathbf{j}}_{s}=-\frac{ie}{m^{*}}(\psi^{*}\nabla\psi-
\psi\nabla\psi^{*}) -\frac{4e^{2}}{m^{*}c}|\psi|^{2}{\mathbf{A}}
\end{equation}

Now let us calculate the OP and the properties of
 the mixed state in the external magnetic field.
It is convenient to  introduce dimensionless variables,
$\psi=\psi_{0} f$, $p_{x,y}=\psi_{0}{\mathcal{P}}_{x,y}$, where
$\psi_{0}=\sqrt{|\alpha (T-T_{c})|/\beta}$,
${\mathbf{r}}=\lambda(T){\mathbf{\rho}}$,
$\lambda(T)^2=\frac{m^{*}c}{16\pi e^{2}\psi_{0}^{2}}$,
${\mathbf{\mathcal{A}}}=\frac{2e}{\hbar c}\xi(T){\mathbf{A}}$,
where $\xi(T)^2=-\frac{\hbar^{2}}{2m^{*}\alpha(T-T_{c})}$ and
${\mathbf{h}}=\frac{2e}{\hbar c}\xi(T)\lambda(T){\mathbf{H}}$.
Separating the phase and the modulus of the OP,
$f=f_{0}\exp{(\phi)}$, and redefining the vector potential the
system of equations is reduced to the equation for the modulus of
the OP $f=f_{0}$ and for the magnetic field
${\mathbf{h}}$\cite{saintjames}:
\begin{equation}
-\frac{\nabla}{\kappa}f_{0}+\frac{1}{f_{0}^{3}}
(\nabla\times{\mathbf{h}})^{2} = f_{0}-f_{0}^{3}
\end{equation}
\begin{equation}
f_{0}^{2}{\mathbf{h}} = \frac{2}{f_{0}}\nabla f_{0} \times
\nabla\times{\mathbf{h}}-\nabla\times \nabla\times{\mathbf{h}},
\end{equation}
where $\kappa = \lambda/\xi$ is Ginburg-Landau parameter.
Amplitude of the p-wave component is expressed in terms of $f_{0}$
and ${\mathbf{h}}$ as:
\begin{eqnarray}
{\mathcal{P}}_{x,y} = \mp\frac{i\eta}{\alpha_{p}\xi}
(\frac{i\nabla}{\kappa}-\frac{curl{\mathbf{h}}}{f_{0}^{2}})_{x,y}
f_{0}
\end{eqnarray}
Therefore the problem is reduced to the standard solution of the
Eqs.(6,7) and then to the calculation of the p-wave amplitudes
applying Eq.(8).

{\it Weak magnetic field $H_{c1}\lesssim H \ll H_{c2}$. Single
vortex.} Solution of Eqs. (6,7) are known and can be written in
the form\cite{saintjames}:
\begin{eqnarray}
f_{0}=c\rho+...\nonumber \\
h=h(0)-c\rho^{2}/2\kappa +...
\end{eqnarray}
where $c$ is the constant of the order of 1. This expansion is
valid near the vortex core $\rho \ll 1/\kappa$. The amplitude of
the p-wave component in that case is:
\begin{equation}
{\mathcal{P}}_{x} = -i{\mathcal{P}}_{y} = \frac{\eta
c}{\alpha_{p}\xi \kappa} (\frac{x-i y}{\rho})
\end{equation}
This effect is consistent with the symmetry arguments and was
briefly discussed in connection with superfluidity of
$^{3}He$\cite{volovik}. Near the vortex core the screening current
is flowing around the vortex as well as the gradient of the
modulus of the OP is finite. As it was shown in Ref.\cite{kab}
currents generate $ip$ secondary OP. On the other hand a real
vector breaks the inversion symmetry and generates the real
$p$-component of the secondary OP \cite{gr}. Far from the vortex
core but in the region where the screening current is strong
$1/\kappa \ll \rho \ll 1$ $f_{0}^{-2}dh/d\rho = -(\kappa
\rho)^{-1}$ and $f_{0}^{2} = 1-(\kappa
\rho)^{-2}$\cite{saintjames} the amplitude of the p-wave component
are
\begin{eqnarray}
{\mathcal{P}}_{x} = -\frac{\eta}{\alpha_{p}\xi \kappa}
\frac{i y}{\rho}\nonumber \\
{\mathcal{P}}_{y} = \frac{\eta }{\alpha_{p}\xi \kappa} \frac{i x}
{\rho}
\end{eqnarray}
This is again consistent with the previous results. Far from the
vortex core the modulus of the OP is almost constant (gradient is
small) and the screening current is strong. Therefore only $ip$
secondary OP survives in that area\cite{kab}.

{\it Strong magnetic field $(H_{c2}-H)/H_{c2} \lesssim 1$}
Magnetic field in that case is determined by the
formula\cite{saintjames}:
\begin{equation}
h=\kappa+\epsilon -\frac{f_{0}^{2}}{2\kappa}
\end{equation}
where $\epsilon=\kappa \frac{H-H_{c2}}{H_{c2}}$.  Spacial
dependence of $f_{0}$ is determined by\cite{saintjames}:
\begin{eqnarray}
f_{0}^{2} &=&|c_{0}|^{2}\sum_{m,n} (-1)^{mn} \exp{(-i\pi
n/2)}\nonumber \\& & \exp{(-\pi(m^{2}+n^{2}-m n)/\sqrt3)} \nonumber \\
& &\exp{(3^{1/4}\pi^{1/2}\kappa i (n x + (2m-n)y/\sqrt{3}))}
\end{eqnarray}
where $|c_{0}|^{2}=(H-H_{c2})/\beta H_{c2}$ and $\beta=1.1596$.
Substituting Eqs.(12,13) to Eq.(8) we obtain
\begin{equation}
{\mathcal{P}}_{x}=-i{\mathcal{P}}_{y}=\frac{\eta}{\alpha_{p}\xi
\kappa} (\frac{\partial f_{0}}{\partial x}+i\frac{\partial f_{0}}
{\partial y}).
\end{equation}

$Weak$ $SOC$. In the strong magnetic field when Zeeman energy is
larger then the energy of the SOC spin decouples from the lattice.
In that case vector functions
$\mathbf{p}_{x}(\mathbf{R}),\mathbf{p}_{y}(\mathbf{R})$ will be
coupled to the magnetic field $\mathbf{H}$(similar effect was
considered in Ref.\cite{sammar}). Therefore linear in gradients
term in the FE has the following form:
\[
 i\eta ' (\nabla \times \mathbf{A})\Bigl
[\psi^{*} (D_{x}\mathbf{p}_{x}-D_{y}\mathbf{p}_{y})-\psi
(D_{x}^{*}\mathbf{p}_{x}^{*}-D_{y}^{*}\mathbf{p}_{y}^{*})\Bigr ].
\]
Here we use cartesian basis in the spin space\cite{varshalovich}
and therefore the triplet OP has zero projection of the spin to
the direction of the magnetic field\cite{lebed}. Minimizing the FE
for variation of the OP $\delta \psi,\delta \mathbf{p}_{x},\delta
\mathbf{p}_{y}$ we obtain the following equation for $\psi$:
\begin{equation}
\frac{\hbar^{2}}{2m}{\mathbf{D}}^{2}\psi+\alpha(T-T_{c})\psi+
\beta |\psi|^{2}\psi-\frac{\eta '^{2}}{\alpha_{p}}
\mathbf{H}\mathbf{D}^{2}(\mathbf{H}\psi) =0,
\end{equation}
where $\mathbf{p}_{x,y}=\pm \frac{i \eta '}{\alpha_{p}}
D_{x,y}(\mathbf{H}\psi)$. If we assume that the magnetic field is
constant and equal to the external magnetic field Eq. (15) is
reduced to Eq. (4) with the field dependent effective mass
$(m^{*}(H))^{-1}=m^{-1}-2\eta '^{2}H^{2}/\alpha_{p}$.
Straightforward variation with respect to vector potential
$\mathbf{A}$ gives the expression for superconducting current
$\mathbf{j}_{s}$:
\begin{eqnarray}
{\mathbf{j}}_{s}&=&-\frac{ie}{m}(\psi^{*}\nabla\psi-
\psi\nabla\psi^{*}) -\frac{4e^{2}}{m^{*}(H)c}
|\psi|^{2}{\mathbf{A}}\nonumber \\
& & +\frac{2 i e \eta'^{2} H}{\alpha_{p}}(\psi^{*}\nabla(H \psi)-
\psi\nabla(H \psi^{*}))\nonumber \\
& & +\frac{\eta'^{2}}{\alpha_{p}}curl
(\psi^{*}\mathbf{D}^{2}(\mathbf{H}\psi) +
\psi\mathbf{D}^{*2}(\mathbf{H}\psi^{*}))
\end{eqnarray}
To simplify this equation let us assume that the magnetic field is a
constant. After simple calculations we get the formula for
$\mathbf{j}_{s}$,
\begin{eqnarray}
{\mathbf{j}}_{s}&=&-\frac{ie}{m^{*}(H)}(\psi^{*}\nabla\psi-
\psi\nabla\psi^{*}) -\frac{4e^{2}}{m^{*}(H)c}
|\psi|^{2}{\mathbf{A}}\nonumber \\& & +
\frac{\eta'^{2}}{\alpha}\nabla \upsilon \times \mathbf{H},
\end{eqnarray}
where $\upsilon = -\frac{2m^{*}(H)}{\hbar^{2}}(\alpha(T-T_{c})
|\psi|^{2}+\frac{\beta}{2}|\psi|^{4})$. It is  seen from this
equation that the correction to the supercurrent is similar to the
ordinary Hall current, where an effective electric field is
proportional to the gradient of the superfluid density,
$n_{s}=|\psi|^{2}$.

When the magnetic field increases further effective mass
$m^{*}(H)$ becomes small and next order terms in the gradients
should be considered in the expansion of the FE density. This is
important for calculation of the upper critical field $H_{c2}$.
For the calculation of $H_{c2}$ we analyze linearized equation
with constant magnetic field directed along $\mathbf{z}$ axis.
Therefore only $z$ component of $\mathbf{p}_{x,y}$ OP is relevant.
We also write  second order gradient terms for p-wave component in
the FE density as $\frac{\hbar^{2}}{2m_{p}}(
|\mathbf{D}p_{x}|^{2}+ |\mathbf{D}p_{y}|^{2})$. Introducing new
dimensionless variables,
$\mathcal{H}=\frac{2\pi\xi_{0}^{2}H}{\Phi_{0}}$,
$\xi_{0}^{2}=\frac{\hbar^{2}}{2m\alpha T_{c}}$, $\Phi_{0}$ is the
flux quanta and  $x \rightarrow x\sqrt{\mathcal{H}}/\xi_{0}$, one
obtains the following set of equations:
\begin{eqnarray}
& &\mathcal{H}(-\psi^{''}+x^{2}\psi)+\tau \psi +\mathcal{H}^{3/2}
\nu(p_{x}^{'}+ixp_{y})=0 \nonumber\\
& &\mathcal{H}(-p_{x}^{''}+x^{2}p_{x})+bp_{x}-\nu c
\mathcal{H}^{3/2}\psi^{'}=0\\
& &\mathcal{H}(-p_{y}^{''}+x^{2}p_{y})+bp_{y}-i\nu c
\mathcal{H}^{3/2}x\psi=0 \nonumber,
\end{eqnarray}
where $\nu=\frac{\eta'\Phi_{0}}{2\pi\xi_{0}^{3}\alpha T_{c}}$,
$c=m_{p}/m \sim 1$ and $b=\alpha_{p}c/\alpha T_{c} \sim 1$,
$\tau=(T-T_{c})/T_{c}$. Here we use Landau gauge
$\mathbf{A}=(0,Hx,0)$ and therefore assume that all functions
depend on x only. Using standard substitution:
\begin{eqnarray}
\psi &=&\sum s_{n}\exp{(-x^{2}/2)} H_{n}(x)\nonumber\\
p_{x,y}&=&\sum q^{(x,y)}_{n}\exp{(-x^{2}/2)} H_{n}(x),
\end{eqnarray}
where $H_{n}(x)$ are hermitian polynomials. After simple
calculations we obtain the following algebraic equations which
determines critical temperature in the magnetic field:
\begin{eqnarray}
&&((2n+1)\mathcal{H}+\tau)-\nu^{2}c \mathcal{H}^{3}
\Bigl[\frac{n+1}{(2n+3)\mathcal{H}+b}+\nonumber\\
&&\frac{n}{(2n-1)\mathcal{H}+b}\Bigr]=0.
\end{eqnarray}
Upper critical field is determined by the highest possible critical
temperature which corresponds to $n=0$ in the Eq. (20). Hence the
critical transition temperature  in the magnetic field is determined
as
\begin{equation}
T_{c}(\mathcal{H})/T_{c}=1-\mathcal{H}+
\frac{\nu^{2}c\mathcal{H}^{3}}{3\mathcal{H}+b}.
\end{equation}
There are a number of interesting observations based on Eq.(21).
The presence of LI increases critical temperature of the
superconductor. This effect is well known\cite{minsam}. More
interestingly, the correction to the upper critical field becomes
nonlinear and leads to the positive curvature of $H_{c2}$ as the
function of temperature: $d^{2}H_{c2}(T)/dT^{2}>0$. Further
increase of the magnetic field may lead to the recovery of
superconductivity in higher fields\cite{recov}. It is also
consistent with the previous results\cite{minsam}, where it was
shown that increase of the critical temperature due to formation
of inhomogeneous helical phase is proportional to $\eta^{2}$.
Therefore, calculated upper critical field should be substantially
larger then paramagnetic or Clogston limit\cite{clogston,lebed}.
Generalizing this result to the weak SOC we can see that the shift
of the critical temperature due to purely paramagnetic effect is
proportional to $\eta'^{2}H^{2}$. This is exactly the results for
$T_{c}(H)$ which follows from Eq.(21). We have to remember that in
high magnetic field higher order gradient terms should be
considered. The higher order orbital effect is proportional to
$\mathcal{H}^{2}$ and reduce the effect of LI. If higher order
orbital effects are strong enough the reentrant superconductivity
in high magnetic field will be suppressed completely.

In Figure 1 we plot the temperature dependence of $H_{c2}$. To
avoid infinite increase of critical temperature in the high field
we take into account higher order gradient terms in the formula
Eq.(21). As a result critical field is determined by the formula:
\begin{equation}
T_{c}(\mathcal{H})/T_{c}=1-\mathcal{H}-\gamma\mathcal{H}^{2}+
\frac{\nu^{2}c\mathcal{H}^{3}}{\gamma_{p}\mathcal{H}^{2}+3\mathcal{H}+b},
\end{equation}
where $\gamma$, $\gamma_{p} \sim 1$ dimensionless constants
describing the effect of the higher order gradient terms in the
Landau expansion. To demonstrate the temperature dependence of the
critical field we choose $\gamma=1/4,b=1,\gamma_{p}=1$ and
$\nu^{2}c$ is changing from 1, to 1.15. The results, presented in
Fig. 1 are restricted by standard range of applicability of
Ginzburg-Landau theory and the limit $T \to 0$ should be
considered only as a qualitative. In that case more accurate
microscopic equation \cite{wert} should be analyzed.
\begin{figure}
\begin{center}
\includegraphics[angle=-0,width=0.45\textwidth]{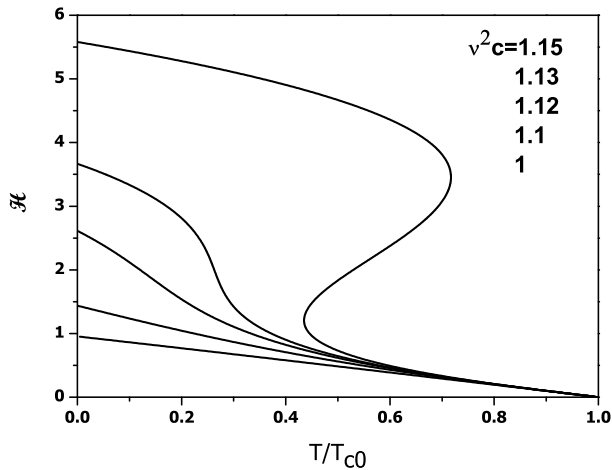}
\vskip -0.5mm \caption{Temperature dependence of the upper
critical field $H_{c2}$, parameter $\nu^{2}c=$1, 1.1, 1.12, 1.13,
1.15}
\end{center}
\end{figure}

To estimate the strength of the effect we refer to Gorkov's
equation formulated for the case of the strong SOC in\cite{kab},
and in the case of paramegnetic effect in\cite{lebed}. It follows
from the analysis in the strong SOC limit \cite{kab}
$\Delta_{p}/\Delta_{d}\approx\Delta_{SO}/E_{f}
\approx\Delta_{SO}\mu_{B}H_{c2}/\Delta_{d}^{2}$, where
$\Delta_{p,d}$ is the $p(d)$ gap, respectively, $\Delta_{SO}$ is
the energy of the SOC, $E_{f}$ is the Fermi energy, and $\mu_{B}$
is the Bohr magneton\cite{err}. From the phenomenological
consideration it follows that $\Delta_{p}/\Delta_{d}\approx \eta/
\xi\alpha_{p}$. Therefore LI are determined by the SOC $\eta/
\xi\alpha_{p}\approx\Delta_{SO}/E_{f}$. In the case of the weak
SOC the estimate reads
$\Delta_{p}/\Delta_{d}\approx\eta'H/\xi\alpha_{p}\approx
\mu_{B}H_{c2}/\Delta_{d}$\cite{lebed}. In that case LI are
determined by by the paramagnetic effect. Therefore at the value
of the field $H\approx H_{c2}\Delta_{SO}/\Delta_{d}$ we expect a
crossover from the strong to the weak SOC limit. The
characteristic value of triplet gap was evaluated in\cite{lebed}
and may be of the order of 0.8 of $\Delta_{d}$ in superconductors
where $H_{c2}$ is close to the paramagnetic limit. As a result we
can estimate the value of the parameter $\nu^{2} c \approx
(H_{c2}/\Delta_{d})^{2} \sim 1$ in Fig. 1. Therefore we believe
that the effect may be observed in materials with weak SO coupling
and with high upper critical field $H_{c2}$, that exceeds
paramagnetic limit. Possible candidates for this effect are
organic superconductors, where highly nonlinear temperature
dependence of the upper critical field is observed \cite{akz}. On
the other hand, the p-wave component of the order parameter in the
mixed state of superconductors may be observed by the direct
vortex imaging with the scanning tunnelling spectroscopy. This
effect should be strong in materials with the strong SOC.

In conclusion we have shown that the additional triplet secondary
OP is generated in the vortex (Abrikosov) phase of
 type II d-wave superconductor. This effect is caused by the
superconducting currents and by the spacial variations of the OP
in the presence of the magnetic field. According to estimates
using the microscopic theory the triplet component may be as large
as 80\% of the primary d-wave component in the field near
$H_{c2}$\cite{lebed}. Nonlinear corrections in the vector
potential to the London equation are derived. The upper critical
field $H_{c2}(T)$ is calculated in the limit of weak SOC. It is
shown that temperature dependence of $H_{c2}(T)$ is highly
nonlinear near the transition to the helical phase.

Enlightening discussions with A. S. Alexandrov, D. Agterberg, A.
G. Lebed, and  V. P. Mineev are highly appreciated. I am also
grateful to G.E. Volovik for drawing my attention to
ref.\cite{volovik} and comments.


\begin{thebibliography}{99}
\bibitem{err} Thereis a missprint in Eq.(9) of Ref.\cite{kab}. The
correct Eq.(9) does not contain $\Delta_{0}/E_{f}$ in the right
hand side.
\end{thebibliography}
\end{document}